# Coordinated Science Laboratory 70[th] Anniversary Symposium: The Future of Computing

*Chairs:* Klara Nahrstedt (UIUC), Naresh Shanbhag (UIUC)

*Technical Program Committee (TPC)* (in alphabetical order): Vikram Adve (UIUC), Nancy Amato (UIUC), Romit Roy Choudhury (UIUC), Carl Gunter (UIUC), Nam Sung Kim (UIUC), Olgica Milenkovic (UIUC), Sayan Mitra (UIUC), Lav Varshney (UIUC), Yurii Vlasov (UIUC)

*Speakers* (in alphabetical order): Sarita Adve (UIUC), Rashid Bashir (UIUC), Andreas Cangellaris (UIUC), James DiCarlo (MIT), Katie Driggs-Campbell (UIUC), Nick Feamster (UC), Mattia Gazzola (UIUC), Karrie Karahalios (UIUC), Sanmi Koyejo (UIUC), Paul Kwiat (UIUC), Bo Li (UIUC), Negar Mehr (UIUC), Ravish Mehra (Facebook Reality Lab), Andrew Miller (UIUC), Daniela Rus (MIT), Alex Schwing (UIUC), Anshumali Shrivastava (Rice U.)

1. Introduction

The Coordinated Science Laboratory (CSL), an Interdisciplinary Research Unit (IRU) in the Grainger College of Engineering at the University of Illinois Urbana-Champaign (UIUC) is a multidisciplinary research institute with a rich history of scientific and engineering advances. It brings together researchers from computing, communication, control, circuits, and beyond. CSL's research portfolio covers the full computing stack, from circuits and high-performance applications to signal processing, machine learning, security and trust, and computing's impact on society and the resulting need for social responsibility. CSL celebrated its 70[th] anniversary in 2021 by hosting the Future of Computing Symposium (FCS) on October 20-21, 2021. In this white paper, we summarize the major technological points, insights, and directions that speakers brought forward during the FCS.

The first session provided an overview of CSL's historical impact on computing. It was followed by a series of sessions that discussed future directions in computing, with an emphasis on emerging research topics and challenges. The sessions' topics included (1) Beyond Traditional Computing, (2) Security and Privacy, (3) AI for Cyber-Physical Systems, (4) Immersive Technologies, and (5) Robotics and Autonomy. Each session was organized in the format of a panel consisting of one external senior researcher and two emerging thought leaders from UIUC. The panel members engaged in extensive discussions on how research in computing, communication, control, and circuits is likely to evolve to advance digital transformation. The discussion brought out the importance of considering not just individual research areas, but innovations at the intersections between research areas and between research efforts and relevant application domains, such as health care, transportation, energy systems, and manufacturing.

Section 2 provides perspectives on CSL's contributions and impact on computing over the last 70 years, some of which were provided by the University, College, and CSL leadership on the first day of the FCS event. Section 3 discusses emerging disruptive technological paradigms comprising "Beyond Traditional Computing," including opportunities at the intersection of brain research and computational science, biological computing, and quantum computing. Section 4 summarizes the "Security and Privacy" session's discussions on privacy and trust in computing, highlighting this area as a major sociotechnical challenge, with deeper discussions on secure learning in adversarial environments and the use of blockchains to enable secure computing platforms. "AI for Cyber-Physical Systems" is discussed in Section 5, which presents challenges to achieving scalable and sustainable AI for everyone, social computing, and true intelligence and enabling therefore comprehensive human-cyber-physical systems (HCPS). The panel on "Immersive Technologies" is summarized in Section 6, and the panel on "Robotics and Autonomy" is summarized in Section 7. Concluding remarks are in Section 8.

## 2. CSL - A Historical Perspective

CSL has a long history of advancing computing, communication, control, and circuits paradigms, methods, and technologies. *Provost Cangellaris* noted that CSL emerged from federal investments following WWII that were intended to improve universities' ability to benefit society and advance prosperity. Since its founding in 1951, CSL has made numerous contributions to advancing computing; many breakthrough ideas and inventions have originated within CSL, and many of its students have gone on to become leaders in academia and industry. Many alumni have acknowledged and reflected on the importance of their time at CSL. CSL was and is at the center of developing strong **collaborative partnerships** reflected in productive centers and institutes. Over the past 70 years, CSL has empowered people to excel by enabling a rich exchange of ideas, in part through the many multidisciplinary collaborations within CSL its research groups, centers, and institutes. In closing, the Provost was pleased to see that CSL, as it looks to the future of computing, is also considering new computing research advances, with due consideration being given to the responsible and ethical development and deployment of artificial intelligence and others.

From The Grainger College of Engineering's viewpoint, *Dean Bashir* stressed the importance of the **multidisciplinary** approach that CSL has fostered over the past 70 years. He observed that researchers from multiple departments, including Aerospace Engineering, Computer Science, Electrical and Computer Engineering, Industrial & Enterprise Systems Engineering, Materials Science and Engineering, Mechanical Science and Engineering, and Physics, among others, have come together in CSL, and continue to do so, to implement transformative ideas. Examples of their achievements include the PLATO system in the 1970s and 1980s, which transformed online education and introduced the first multiplayer video game, called Empire; the electric vacuum gyroscope in the 1950s and 1960s, which has since been deployed into the deepest depths (submarines) and up to the greatest heights (spacecraft); and multiple generations of the ILLIAC computer systems in the 1950s to 1970s. This rich tradition of multidisciplinary achievements is the foundation upon which CSL will build the next generation of future transformative ideas, moonshot goals, and transformational research at the intersection of computing, communication, control, and circuits. Furthermore, the centers and institutes within CSL (e.g., the Center for Cognitive Computer Systems Research ($C^3$SR)[1], the IBM-Illinois Discovery Accelerator Institute[2], the Center for Networked Intelligent Components and Environments (C-NICE)[3], IoBT REIGN[4], the Center for Advanced Electronics through Machine Learning (CAEML)[5], and the Center for Autonomy[6]), fueled by the creativity and drive of faculty, students, and staff, represent intellectual platforms for the next generation of multidisciplinary discoveries, innovations, and accelerators of future computing research and technologies.

---

[1] https://www.c3sr.com/
[2] https://discoveryacceleratorinstitute.grainger.illinois.edu/
[3] https://cnice.csl.illinois.edu/
[4] https://iobt.illinois.edu/
[5] https://caeml.illinois.edu/
[6] https://autonomy.illinois.edu/

Collaborative partnerships with other academic institutions, government agencies, and industry, and multidisciplinary collaborations among faculty and students spanning multiple departments, have been hallmarks of CSL from its very beginning. Another other important hallmark of CSL was, is, and continues to be, its **integrative approach to theory and practice**, as noted by the *Director of CSL, Klara Nahrstedt*. CSL researchers' theoretical results have always informed the building of novel computing systems, and vice versa. This started with the development of the first digital computer to have a compiler (ORDVAC, in 1952) and the ILLIAC II computer system (1952), which was used to calculate Sputnik's trajectory. This integrative approach continued with the development of the PLATO platform as the first computer-assisted instructional system. PLATO had a fundamental and transformative impact on multiplayer games, animation, touchscreens, display graphics, and development of online communities. CSL's integrative approach continues today with advances in theory and experimental systems spanning areas such as computational theory, robotics and control, the speed limit of microprocessors, CAD tools and algorithms, signal reconstruction to improve signal fidelity, acoustics, computer vision, omni-focus cameras, reliable and trusted computing, computer networks, network delay calculus, scheduling and routing of packets, optical networks, scaling laws in wireless networks, and many other fundamental and translational aspects of current and future computing environments.

### 3. Beyond Traditional Computing

Alternative computing paradigms, such as neuroscience-based computing, biological computing, and quantum computing, are increasingly being advanced as possible alternatives or complements to traditional computing. As we reach the limits of Moore's law, silicon-based von Neumann computing cannot be the only basis for the next generation of computational systems. Future computing paradigms must consider hybrid approaches, and, perhaps, a potential transition to entirely new computing paradigms, materials, and technologies.

#### 3.1. Neuroscience-inspired Computing

The human brain is widely regarded as the ultimate computing engine, with extremely high energy efficiency, reliability, and learning and cognitive capabilities. The aspiration to design computing platforms with attributes approaching those of the brain is widespread and longstanding. Neuroscience-inspired computing relies greatly on understanding the human brain and leveraging that understanding to design computing platforms that approach the brain's efficiency, functionality, and robustness. This area lies at the intersection of brain/cognitive sciences and computing. A more mature manifestation of neuroscience-inspired computing is *neuromorphic computing*, which had its origins in the pioneering work of Carver Mead and his collaborators in the 1980s. Although much progress has been made since then, neuromorphic systems have yet to demonstrate the cognitive functionality of mainstream AI methods (e.g., deep nets), or energy efficiency approaching that of the brain. A key reason for this situation is that our understanding of the human brain is still in its infancy.

*Dr. James DiCarlo (MIT)* discussed the importance of "reverse engineering" the brain to understand its capabilities. One approach is to measure the dynamics of brain components and map the resulting topologies (i.e., connected neurons) into engineering formalisms, such as the predictive building of systems, to construct scientific hypotheses regarding brain functionality. The goal is to create a "virtuous cycle" between brain understanding and computing that can advance our understanding of human intelligence, especially visual intelligence.

Dr. DiCarlo and his group conducted a variety of visual brain experiments to study how long it takes the brain to recognize an object (e.g., "is a car there?"). Using electrophysiological recordings from animals and neuroimaging techniques with animal and human subjects, Dr. DiCarlo is studying the patterns of brain activity that underlie our ability to recognize objects visually. He and his collaborators identified and analyzed the full ventral visual stream, from the rapid movement of eyes and constraints on user behavior to the brain's neural activity. They have been able to determine that core object recognition takes approximately 200 ms. The full path identification required analysis of each brain component participating in visual object recognition. The group collected brain-related data at every step, and reported, for example, that visual data are transformed into patterns of neural activity that are transmitted into the brain and eventually arrive at the inferior temporal (IT) cortex 100 ms later. These patterns are of great interest, since they are cognitively relevant and represent codes that can be mapped into artificial neural network (ANN) models within engineering/computing systems. As we improve our understanding of what is happening in the brain, and especially in the IT cortex, we can improve on the ANN models, scientific hypotheses, and alignment with observed behaviors. Over five years, Dr. DiCarlo and his group evaluated hundreds of computational ANN models with respect to their alignment with the brain. They have built a platform called Brain-Score[7] that includes brain alignment benchmarks.

Although our ability to model visual intelligence is advancing, new generations of ANN models that improve on the current ANNs and behavioral models are needed to align the brain's ventral stream with future computational models. Scientific understanding of natural intelligence and in silico intelligence are thus interlocked aspects of a single grand challenge.

### 3.2. Biological Computing

In a second alternative computing paradigm - biological computing - computation occurs through use of biological elements. It is thus different from the bioinspired computing paradigm. For example, proteins are directly used to compute binary operations or other digital mathematical operations. Cellular substrates are also now of interest, spurred by biological studies, on a range of animal models, targeted at understanding how neurons grow, how they are used, and how one can control a biological system to the point that it can become part of a hybrid computing ecosystem. For example, a nematode (a kind of non-segmented worm) has only about a hundred neurons, but it can crawl, explore, and look for food or mates. Cognitive tests have shown that a nematode is capable of rudimentary forms of learning and awareness and has a range of cognitive functions and behaviors that our traditional computing machines struggle to replicate under similar complexity constraints. Another example is fruit flies, which have 100,000 neurons and the fastest visual system in the world. Bees have one million neurons and can form social structures, divide labor among themselves, and engage in complex social behaviors.

*Dr. Mattia Gazzola (UIUC)* and his group study biological systems and machines. Biological machines are analog and stochastic, and hence are inherently primed for the kind of free-form computation that is needed for learning, awareness, association, and synthesis of capabilities. However, the current "finite-state" computing machines and models are digital, deterministic, and rigid. Dr. Gazzola therefore asks questions such as, can we understand the algorithmic basis of behaviors? And can we outfit traditional computing with cognitive primitives? It is important to stress that computing and the body are not necessarily separate in biological creatures. Some animals can use their bodies to compute various functions. Can we achieve similar capabilities in future computing platforms? One potential answer is the development of an in vitro bio-hybrid system as a behavioral and computing discovery platform. An example of that approach is the Mind in Vitro project[8] about computing with living neurons.

---
[7] https://www.brain-score.org/

An interesting case study that Dr. Gazzola is investigating is the replacement of hardwired connectivity in a Braitenberg vehicle[9] with neurons. A Braitenberg vehicle agent can autonomously move using sensors that measure stimuli and are connected to wheels that serve as actuators/effectors. In a traditional Braitenberg vehicle, one can create the connectivity between sensors and actuators to enable generation of a variety of behaviors, but once the connectivity has been set, it is unchangeable. A goal of the group's current research is to render that mechanical connectivity malleable by leveraging neural plasticity. One would not reverse-engineer the brain, but instead enable growth of neurons with specific properties, adding human design to control connectivity and flow of information to carry out computations. Although further neuromuscular integration and functional plug-and-play 3D neural tissue topics are being explored, many outstanding issues must be studied, including ones related to computer architectures, programming languages, 3D electronics for I/O interfaces, and many other computing concepts, to make in vitro bio-hybrid systems ready for next-generation computing applications.

### 3.3. Quantum Computing

Quantum computing, one of the most exciting yet challenging topics in quantum information science (QIS), is a method of computation that utilizes the collective properties of quantum states, such as superposition, interference, and entanglement, to perform computation. Quantum computing has seen significant growth over the last several years. *Dr. Paul Kwiat (UIUC)* discussed the major differences between the classical "bit" in traditional computing settings and the basic quantum bit ("qubit"). Classical bits are binary, 0 or 1; they are stable and can be copied (cloned). Qubits, on the other hand, can be in states that represent 0 or 1, or in any arbitrary superposition of these states; there is no way to clone qubits. The Bloch sphere is a geometrical representation of the quantum states of a two-level quantum mechanical system (qubit), with pure states corresponding to points on the surface of the sphere. A basic quantum circuit, called a *quantum gate*, can be represented by rotating the sphere about any axis by any angle (*much* richer in terms of states than a bit flip with two states 0 or 1, the only single-bit classical gate); unlike many classical logic gates, quantum gates are necessarily reversible. The heart of the differences between traditional and quantum computing is quantum entanglement, a physical phenomenon that occurs when a group of particles are generated or interact in such a way that the quantum state of each particle of the group cannot be described independently from the state of others. Hence, entanglement and the resulting scaling of correlated states are key to the power of quantum computing.

The main question addressed was, what is quantum computing good for? Quantum computing algorithms have been developed that are relevant for many computing applications. For example, Shor's algorithm for factoring numbers into prime constituents could undermine the security of most of the currently used public key cryptography protocols. Grover's quantum search algorithm can find a marked element in an unsorted database with N elements, and was proven to be optimal, with a sqrt(N) speedup over any classical search. The HHL (Harrow Hassidim Lloyd) quantum algorithm can efficiently solve linear systems of equations.

---

[8] https://miv.illinois.edu

[9] https://en.wikipedia.org/wiki/Braitenberg_vehicle

However, many grand challenges remain. Many algorithms have been discovered[10], but an implementation for them is still lacking. For example, quantum algorithms that can find quantum ground states – a central task for simulating new materials – exist, but it is currently impractical to realize them for all but the simplest systems. Other areas to be developed include distributed quantum computing and hybrid quantum-classical computing. UIUC researchers, including Dr. Kwiat, are leaders in the quantum information revolution, exploring many of these challenges within the Illinois Quantum Information Science and Technology Center (IQUIST), the Hybrid Quantum Architectures and Networks (HQAN) NSF Quantum Leap Challenge Institute, and Q-NEXT, a large DOE-funded QIS research center.

### 3.4. Panel Discussion

**Neuroscience-inspired computing** discussion focused on (1) determining the accuracy of brain measurements, (2) whether spiking patterns provide an accurate measure of brain functions, and (3) how training of ANNs and deep nets can be related to observations of training and learning in the brain.

- **Accuracy of Brain Measurements**: Quantification of accuracy requires a well-defined set of measurements. That was the reason for the creation of the Brain-Score platform. The platform demonstrates that different models yield different accuracies, ranging between 20% and 80%. However, many measurements are not included in the platform.
- **Spiking Patterns in Brain Measurements**: Brain measurements have predominantly concentrated on neural spiking patterns because these are the only known information mechanisms that can operate at the speed needed to explain the brain's visual recognition tasks (i.e., latent variable inference tasks).
- **ANN/Brain Training**: Training of artificial deep networks from different starting points (random initialization) results in different intermediate representations once the non-convex optimization process was completed. It remains to be determined whether that is the case in the human brain too. Different people and monkeys have different brain representations, but the representations also have similarities. The questions pertain to degree and specific tasks. However, ANNs and brain activities for a visual object recognition task perform similarly even when different initial starting points for network training are utilized (in training on the same task).

The **biological computing** discussion covered topics including (1) DNA/molecule-level vs. cell-level computing, (2) a new theory for biological computing, and (3) energetics in biological computing.

- **DNA vs. Cell-level Computing**: There was significant early interest in molecular-level computing, but interest has faded. While there have been advances in extremely parallel in-memory computing with DNA substrates, the time it takes to complete a computation using DNA-level computing is relatively long. Cell-based systems have a natural ability to interface with sensors and cells. Such systems can do memory, computing, and sensing in one, resulting in higher computational speeds. The speed of cell-level computing is determined by the speed of the chemical reactions.

---

[10] See http://quantumalgorithmzoo.org.

- **New Theory**: It is important to have some theory to guide exploration in biological computing. One may ask, when will we stop using words like "software"? There is an assumption that new forms of computing will have "software" and "hardware," but what if we just have algorithms that work like the brain?
- **Energetics**: Neuromorphic computing has been looking at spiking as a potential source of energy efficiency, but it turns out that spiking is not the place to look for energy efficiency. In the brain, spiking serves to transmit signals over relatively long distances. In biological computing, spiking is important as well; the energies expended in neuromorphic computing and biological computing are comparable at the neuron level.

The **quantum computing** discussion included questions on (1) whether we need a new theory to orchestrate classical and quantum computing, (2) what new software paradigms are needed for quantum computing, and (3) whether we will see large-scale cloud computing integrated with quantum computing in a huge cold unit.

- **New Theory**: Recent results, e.g., by Fred Chong from the University of Chicago, show significant benefits in physical-level optimization. Thus, it may not be a good idea for quantum to abstract away the specifics of a particular physical implementation – the precise physical platform can influence the optimal architecture and encoding.
- **New Software Paradigms**: Since a variety of quantum computing technologies are emerging, companies are working on translators for software, e.g., software translation from a Microsoft quantum computer to an IBM quantum computer. Companies also have their own coding algorithms, e.g., error correction algorithms. At UIUC, Brian Doolittle, a Ph.D. student of Prof. Eric Chitambar, developed a quantum software suite[11].
- **Integration**: Some quantum computing platforms can work in room-temperature environments. Thus, it may not be necessary to have quantum computers and cloud computers integrated in a huge cold unit. Regardless, most universities will likely not have a large-scale quantum computer onsite. They will instead run quantum-based programs on a remote quantum processor, accessed via a quantum network.

4. **Security and Privacy**

Security is an important part of all computing systems around us and must be designed in as an integral part of any computing platform or infrastructure. The scale and complexity of these platforms and infrastructures, in terms of the diversity of users, devices, networks, software, hardware, and application services, are increasing. We are therefore facing new security and trust challenges. Future computing grand challenges in security are daunting given the need (1) to consider trusted platforms in diverse social environments, (2) to secure new machine-learning-based workloads, and (3) to develop secure computing platforms that will support cryptocurrency, blockchains, and other new digital transformation paradigms.

4.1. **Trusted Networks in Diverse Social Environments**

Having high-speed, affordable, and reliable Internet access is of crucial importance since it affects many aspects of our everyday lives. We rely on such access for remote education and learning, for remote healthcare and tele-medicine, and for the opportunity to work and be entertained from home. This reliance has, of course, been amplified by the COVID-19 pandemic. *Dr. Nick Feamster (University of Chicago)* presented the existence of an **urban digital divide** as a major problem when it comes to Internet networks and access. Cities have major Internet access inequities. Parts of cities face difficulties in accessing digital services, and Internet service providers (ISPs) have little or no data even to document this lack of access to digital services.

---
[11] https://csl.illinois.edu/news/42559

ISPs would greatly benefit from data/information about application performance (e.g., what people's Zoom experience was like), fine-grained measurements of network traffic, and information on the downstream effects of limited bandwidth in communities to determine how to provide high-quality Internet network access to deprived communities. For example, after the COVID-19 pandemic began, the dramatic increase in the use of teleconference tools like Zoom led to traffic volume increases at Internet interconnection points and a major decrease in available network bandwidth for some communities, which experienced higher latency and lower throughput. Having information about the internal status of the Internet can help ISPs add more bandwidth capacity and improve latency and throughput to communities.

In some cases, ISPs have had difficulty securing such status information because of a **lack of trust** in economically-disadvantaged communities. One could learn about Internet network quality, for example, if one could put a computer in someone's home. But some people will not let ISP providers come and install "boxes" in their homes for fear that spyware could be present and violate their privacy.

Several methods can be used by ISPs to address privacy concerns. They include measuring encrypted traffic at routers, conducting passive measurements, and using AI-based inference on Wi-Fi networks, applications, and activity measurements. Passive measurements can yield important insights on average network utilization, recovery times following transient drops in bandwidth, differences in proprietary congestion control, and how participants' video layout affects their own network utilization and that of others.

However, these solutions still do not result in a gain of trust in underserved communities. Trust is needed to enable passive Internet networking measurements at scale. Achieving **privacy** and securing the **trust of people** is a major sociotechnical challenge. Another important technological challenge is that of ensuring the physical and cyber security of the large-scale distributed measurement systems that ISPs would need to deploy. Questions regarding central control, authentication, deployment of software (that might look like a botnet or cryptominer) must be answered in conjunction with consideration of privacy and trust. Distributed measurement infrastructure represents a large attack surface that must be secured.

### 4.2. Secure Learning in Adversarial Environments

Adversarial learning and the ability to deploy secure learning algorithms in adversarial environments were the second security topic covered during this session. Machine learning (ML) is becoming ubiquitous in many computational and communication digital services. We are seeing increased numbers of attacks on machine learning algorithms, ranging from **poisoning of ML models** and their parameters to **privacy violations and biasing** of ML models to distort data, resulting in **incorrect ML models**. *Dr. Bo Li (UIUC)* investigates the design of robust and certifiable, private, and explainable ML paradigms for real-world applications. She discussed how adversarial attacks on ML models can happen and how one can defend against them.

An example of a simple adversarial attack could be simply **putting some optimized stickers on a traffic sign.** The sign might remain readable to a human but be incorrectly recognized by an automated ML-based system. For example, a stop sign, modified with stickers, might be interpreted by an autonomous vehicle to say, "Speed Limit 45" instead of "STOP." That would, of course, be extremely dangerous.

To defend against adversarial machine learning attacks, multiple solutions must be deployed. First, companies must **detect and disclose the vulnerabilities they have identified**, and second, they must work to **prevent** such vulnerabilities. One direction being explored is "**certified robustness**" and the development of frameworks to certify the correctness of provided ML models.

The certified robustness approach still has a long way to go, since we must find answers to many questions, such as, how do we guarantee overall safety? What are the practical constraints for attackers when handling ML pipelines? How can we integrate human knowledge and cryptography techniques to improve ML robustness? How do we identify vulnerable components and data within an ML pipeline? What does privacy-preserving learning look like?

### 4.3. Secure Platforms for Blockchains

Blockchains have been a hot topic in recent years. Blockchain technology allows for separation (disintermediation) of trust between the application developer and the blockchain platform, which runs a smart contract program. *Dr. Andrew Miller (UIUC)* studies blockchains, and he discussed the opportunities and challenges of blockchain technology during his presentation.

The **disintermediation between the application developer and blockchain platform** can lower barriers to entry for emerging service or product providers. Customers can more readily trust the software services of an application developer, even if the developer doesn't have the kind of established reputation that longstanding providers may have won. The reason is that use of blockchain makes it easier for a customer to verify that the software services being provided are running as they should. For example, a customer does not have to blindly trust a startup company (i.e., service provider) to take care of the customer's funds. The customer can look at the company's code or hire an auditor to review it. Blockchain is being widely used, for example, in the cryptocurrency domain, and there have been a lot of venture capital investments in this area (over $5 billion quarterly in investments in startups).

Blockchain is, however, not well suited to addressing **privacy**; it will be necessary to cope with the privacy issue over the next decade. Blockchain uses replication, which means that every user of the blockchain can fetch a copy of all the on-chain data. This is especially tricky when one deals with sensitive data and excessive access. For example, what prevents a service provider who collects customers' location data from using it for nefarious purposes (say, finding out who attended a protest event), instead of legitimate ones (say, tracking COVID-19 super-spreader events)? To solve this blockchain privacy challenge, a robust roadmap must be developed to secure computing for blockchain, including consideration of the role of cryptography and secure hardware. With respect to **cryptography**, blockchain technology will require **multiparty secure computation**, i.e., it will be necessary to store sensitive data in a secret-shared form, e.g., three out of four servers must agree before data can be accessed. With respect to secure hardware, blockchain technology must be based on secure hardware enclaves such as TEEs (trusted execution environments). A TEE computes on sensitive data only if the user has specifically authorized that query. Secure hardware enclaves enable remote attestation of data, audit logs of queries, and storage of sensitive data that are encrypted and decrypted only within secure tamper-resistant enclaves.

### 4.4. Panel Discussion

The panel discussion focused on four topics: (1) clean-slate design, (2) certification, (3) disintermediation and scale, and (4) secure hardware.

**Clean-Slate Design for Secure and Trusted Platforms**: It is difficult to envisage developing and then implementing a clean-slate design for trusted networks and network measurements, since so much infrastructure is already in place. Internet service providers (ISPs) face other fundamental issues as well. For example, how does an ISP prove to a customer that it does not know something about the customer and/or cannot infer something about the customer from the data it already collected? The issue comes down to, how can we make **privacy-preserving measurements that hold up to audits**?

In ML systems, clean-slate security and privacy design is difficult as well. It is impossible to secure a system against an extremely powerful attacker. One must be realistic about ML-based approaches and their limitations. Given those limitations, questions that must be addressed include, what kinds of guarantees do we want for **accuracy, speed, privacy, and security**? What are the priorities in achieving security and privacy?

**Certification of ML Models**: Companies are training ML models for millions and billions of parameters and offering trained classifiers. This can be advantageous to users since they do not have to train the ML models; the users can immediately use the provided trained classifiers for their classification and prediction tasks. A fundamental question naturally presents itself: how can companies provide verifiable certifications as to the validity of the trained classifiers? **Credit scores** may be the answer. For example, in facial recognition, some companies provide guarantees on how well their AI/ML models recognize faces by quantifying the facial recognition in a credit-score-like format. Mature products and scores that people can readily and reliably depend on are not yet here.

**Disintermediation and Scale of Blockchains**: Blockchains allow disintermediation among the application developer, platform, and customers. There is perhaps no better-known application of this disintermediation property than cryptocurrency. While there are crypto-optimists who believe that blockchain is the future, there are also crypto-pessimists. For example, in the music domain, a musician might get a tiny payment whenever a customer listens to or uses his or her song. However, to succeed, the song must be available on popular platforms, which are the intermediation parties. Hence, the use of cryptocurrency does not eliminate the need for powerful intermediaries, such as these popular music platforms, to reach a sufficiently large customer base. It is not clear which domains might benefit from blockchains and disintermediation, e.g., transaction payments with cryptocurrency without a third-party service provider (e.g., a digital wallet), and which domains might benefit from intermediaries, e.g., transaction payments with cryptocurrency with a third-party service provider (e.g., a music platform). The answer lies in understanding economies of scale and the benefits of having large, central intermediaries as market actors. If one needs robust database services, for example, customers may rely strongly on, and benefit from, the reputation and presence of centralized, established companies such as Oracle. Blockchain technology does not remove the advantages of such well-established companies. It can, however, level the playing field for startup companies. Startups can hope to provide their services on a blockchain platform with less overhead, since a blockchain platform can allow customers to audit the services of a less well-known, or even unknown, company.

**Secure Hardware:** Current **nanoscale circuits** are replete with **noise** and variations of noise (scrambled bits). One might ask whether the current secure hardware could leverage this noise to enhance robustness, security, or privacy. That question has different answers depending on the type of noise present and whether it has the necessary characteristics.

At the chip level, one could consider using **manufacturing noise** during the fabrication of chips/devices to create unique cryptographic keys, since the noise can be a basis for **physical unclonable functions (PFUs)**[12]. The manufacturing/fabrication process of chips makes the device/chip noise unique. One could use the unique features of the chip noise as a key that does not have to be stored in memory, but instead can be recreated whenever the computing hardware needs it. Nanoscale circuit noise is random, however, so the viability of harnessing nanoscale attributes for security and privacy is still an open question.

For ML applications, **computational hardware noise** can be beneficial for robustness certification. This approach requires the ability to analyze and **quantify hardware noise**.

For network infrastructures, **network hardware noise** is important for the robustness of packet transmission models. One can use the noise during the training of packet transmission models, so that if network hardware attacks do happen, network connections can rely on packet transmission models trained under noisy network conditions.

**Energy-efficient hardware** is even noisier, and raises another question: what is the **effort and overhead** we pay for a more robust (secure and privacy-preserving) system in terms of energy and latency? Making the optimal trade-offs between security/privacy and energy/latency is a major challenge given that security and privacy increase the energy usage and latency of systems. It is not quite clear how to narrow down and handle this trade-off in ML systems.

Other concerns related to security in hardware accelerators were discussed. Hardware improvements for **secure computation** are being made rapidly, including improvements to **hardware accelerators**. In general computation, there have been robust performance improvements in GPUs. In multi-party computation, however, GPUs are not the most efficient hardware components. There is a need for high-performance cryptography co-processors, especially in cloud settings.

5. **AI for Cyber-Physical Systems**

Artificial intelligence (AI) and associated machine learning approaches are increasingly becoming integral parts of cyber-physical systems, such as those used in digital manufacturing, agriculture, transportation, and social systems. They are of particular importance in domains in which cyber systems are augmenting the human (social) and physical world via collection, analysis, classification, prediction, actuation, and task automation and ultimately enabling human-cyber-physical systems. In this session, three important topics were discussed. (1) Do we need to process all components (e.g., neurons) in our learning processes to enable scalable and sustainable AI for everyone? (2) Are biases in our social AI-driven, computer-mediated systems diminishing the desired AI return? And (3), how do we get from intelligence augmentation to true artificial intelligence?

---

[12] https://en.wikipedia.org/wiki/Physical_unclonable_function

### 5.1. Scalable and Sustainable AI for Everyone

AI-based deep learning models are evolving rapidly to have ever greater speed and performance accuracy. They are incorporating larger and larger models for decision-making. For example, Google, in 2017 and 2020 (the GShard[13] system), reported models with millions to billions of parameters for translating one hundred different languages into English. Another major development in AI-based models is the replacement of older RNNs (recursive neural networks), LSTMs (long short-term memory networks), and CNNs (convolutional neural networks) with attention and/or transformer models. The ultimate challenge and goal in AI is true artificial intelligence, but in the near term, the objective is to find models that enable **fast training** and maintain good accuracy. The search for correct and cheaper models is currently done using trial-and-error approaches. This practice is neither scalable nor sustainable, for two reasons: (1) the computer power needed to advance AI algorithms doubles every 3.5 months, and the benefits of AI will soon be surpassed by its cost; and (2) advances in AI come with a huge carbon footprint. Current approaches to addressing these concerns are focused on the use of specialized computing hardware, such as GPUs, TensorFlow, and FPGAs. However, these approaches are less than optimal, since the specialized hardware is hard to virtualize.

*Dr. Anshumali Shrivastava (Rice University)* is working on the scalability and sustainability of AI deep learning models by considering alternatives to the **suboptimal back-propagation algorithm** that is part of current AI algorithms and was first introduced by Rumelhart, Hinton, and Williams in 1986[14]. The back-propagation algorithm is used in layered neural networks to fine-tune the weights of a neural network based on the error rates obtained in previous epochs (iterations). Proper tuning of weights reduces error rates and makes the model more useful, since it increases its generality. But the back-propagation algorithm is wasteful because it computes all neuronal activations, which is not always necessary. If one could pick the best neurons for updates, the savings would be tremendous.

That kind of savings is possible if we organize neurons into **hash tables**, as is done within the SLIDE system. One can then apply joint adaptive sampling and **locality-sensitive hashing (LSH)** within the **deep learning** framework. This **LSH + deep learning** framework builds hash tables by processing the weights of the hidden layers and places the active hidden layer's hash table entries into the memory using fingerprints, so that when a query arrives, the framework performs forward- and back-propagation on the active set of nodes. The framework also makes it possible to update the hash tables by rehashing the updated node weights. The resulting overall computation is significantly reduced, since the computation complexity is on the order of the number of active neurons, and not the total number of neurons. This kind of computational efficiency is like that seen in biological systems such as the brain, since the approach utilizes sparse coding.

There are additional directions for research with respect to optimization for scale and sustainability when deploying these LSH + AI-based algorithms. These directions include the utilization of parallel cores, the reduction of memory latency, the reduction of cost while achieving software and hardware acceleration, the use of convergence when conducting heuristic sampling, and many others.

---

[13] https://arxiv.org/pdf/2006.16668.pdf
[14] https://www.nature.com/articles/323533a0

## 5.2. Biases in AI-based Frameworks

The widespread deployment of computer-mediated social platforms has given rise to many challenges and avenues of research in what is called *social computing*. Early examples of computer-mediated systems, such as PLATO, enabled multiplayer gaming (e.g., PLATO's Empire game). Now, advanced computer-mediated systems are becoming widespread social platforms. While these social platforms typically execute some form of content moderation, evidence has increasingly shown that these platforms use algorithms that exhibit **biases**. For example, women do not see some job ads; minorities are negatively affected by healthcare algorithms; and individuals can be falsely arrested based on incorrectly trained and implemented facial recognition platforms.

*Dr. Karrie Karahalios (UIUC)* studies social platforms and the AI-based algorithms, deployed within these platforms, that result in biases. She noted that **algorithmic biases** are widespread. These biases can be seen, for example, in the housing domain. During an **audit**, it was shown by researchers that potential real-estate buyers from black, Hispanic, and Asian populations were shown fewer homes than white buyers were. Alteration of users' social media feeds without their knowledge is another issue. The research system, FeedVis[15], curates news feeds so that people see alternative realities without being aware of the curation. Researchers are still studying the impact of such curated news feeds on users.

Many challenging questions on bias in AI systems remain open. They include questions on whether to reveal the existence of biases in algorithms, how to determine what the user experience should be, how to give users a say in this experience, how to optimally design the user experience with respect to machine learning, and how to allow users to specify bias they detected using participatory design. The possibility of giving users the capability to report that they detected bias in an AI system involves issues that raise numerous research questions with respect to representation, argumentation, cultural competence, comparison, and access. Other questions include the development of control panel theories for controlling settings in social platforms and their algorithms, messaging, explainability of user interfaces for awareness, contestability, and regulation.

## 5.3. From Intelligence Augmentation to Artificial Intelligence

We already have intelligence augmentation that significantly impacts our lives and our society. Going forward, these intelligence augmentation systems will become increasingly "influential" and/or capable of enabling actors to exert influence. It is therefore important to carefully study these systems, and their influence, long before we reach anything close to true artificial intelligence, according to *Dr. Alex Schwing (UIUC)*.

Early forms of "intelligence augmentation" included libraries and encyclopedias. They are slow to search, collect general knowledge, and are passive, i.e., a library or encyclopedia doesn't decide on anything. Over the last 30 years, the Internet and other forms of communication have allowed us to search "intelligence augmentations" faster and expand them using much more data to become personalized. Notably, these "intelligence augmentations" remain largely passive. As an example, think about credit card companies that attach scores to their customers. Those systems can be searched quickly and are personalized, but remain passive, because an employee makes the decision.

---

[15] https://dl.acm.org/doi/10.1145/2685553.2702690

More recent emerging forms of intelligence augmentation, based on **machine learning** (ML), conduct searches quickly, are personalized, and actively make decisions. For example, AI systems can advise a student on what to learn, or tell a farmer what to plant. In banking, AI systems are contributing to loan decisions (if not making loan decisions) based on digital social profiles. Job applications are being screened by algorithms that choose which ones to flag for further review. News feeds are being curated according to engagement time (creating possibly unwanted echo chambers rather than encouraging critical study).

While these emerging systems of intelligence augmentation promise, e.g., increased productivity, improved reach to a particular audience, or a healthier lifestyle, they are also increasingly complex, and the reasons for their recommendations are "murkier." Even more importantly, these systems are being developed at an increasingly fast pace because ML technology is more readily available, a testament to the AI/ML community. However, the foundational understanding of many of these real-world systems is often lacking because necessary studies are time-consuming and very expensive, and often involve large human-subject studies.

It is therefore worthwhile to discuss whether an approach like the FDA's drug review process needs to be established for AI applications. Admittedly, such a review process might significantly slow the progress of adoption. To avoid that, the future of computing may require more effective automated mechanisms for understanding the AI/ML systems we are building. Such mechanisms are in their infancy and unfortunately are not (yet) advancing as quickly as the developed intelligence augmentation systems.

Importantly, all these advances are taking place in the context of our society. Studies of these AI/ML systems, whether automated or via a particular entity (e.g., human/review committee), are societally significant. They must be comprehensive and able to answer: what should be the role of intelligence augmentation in society moving forward?

### 5.4. Panel Discussion

The panel discussed three themes: (1) trade-offs between the number of activations and consumption of storage, energy, and other resources; (2) biases and ethical considerations related to data sampling, model training, and learning; and (3) challenges of intelligence augmentation.

**Trade-offs in Reducing the Number of Activations**: It is not clear whether there are always clear savings from reducing the number of activations in neural networks, because the process for doing so introduces other overheads. Energy efficiency is one area of concern. For example, one must take the overhead of maintaining hash tables into account. The SLIDE system provides clear speed-ups of neural network training. Reading data from memory, however, is energy consuming. Trade-offs must be studied when considering new neural network approaches.

**Biases and Ethical Considerations for Data**: There is a push to have an "**AI Bill of Rights**" that would be a living document that incorporates community beliefs. The bioethics community deals with biases and ethical issues in its domain in part by using a similar approach. Another question is whether unbiased datasets even exist. If we do not have access to datasets produced by and for diverse groups of people, biases occur. There have been cases in which companies trained AI systems on U.S. or European datasets (thus creating unintentional but significant biases), and then deployed them in Asian markets. Indeed, the issue of biases in datasets is a major challenge. Before using a dataset, one must carefully consider how and where the data are collected, and whether technologies that have been tested on one population might be presented to another. A good example of problematic datasets is in genomics research. Only people who are relatively well-off financially can pay for obtaining their genomic profiles. As a result, the available data are skewed towards that population, and cannot be used for accurate analysis of rare diseases and diseases that affect specific populations.

**Challenges of Intelligence Augmentation**: ML/AI systems will help make many processes more efficient. Identifying biases, establishing the **verifiability of biases**, and correctly addressing such biases are challenging areas of research in this domain. We must ask, and find ways to answer, questions such as, can we make a **system fair** even if it is based on biased input? Furthermore, there are many tasks that only **humans** can do. For example, companies that aim to automate the review and moderation of content are not doing nearly enough, and their automated reviewing is often not as good as it needs to be. Until significant advances are made, human resources likely must be kept in the loop.

## 6. Immersive Technologies

Immersive computing promises to enable humans to communicate and interact in a natural way in shared immersive spaces. Many research efforts are underway to provide solutions for immersive audiovisual devices and immersive extended reality environments in which humans can seamlessly and naturally interact and communicate with each other. In this session, topics such as the role of immersive audio in the future of human communication, the challenges of end-to-end hardware-software co-design for immersive systems, and brain interfaces for better immersion of humans in virtual environments were discussed.

### 6.1. Immersive Audio and the Future of Human Communication

Human communication would be much more difficult without sound, which is fundamental to both speech and music and thus plays an indispensable role in human communication. Sound can carry emotional richness unlike that of any other sensory experience. As Helen Keller, the deaf and blind American author, said, "Blindness separates people from things; deafness separates people from people."

*Dr. Ravish Mehra (Facebook Reality Lab)* discussed the status of human-human computer-mediated communications and the many challenges that must be addressed to achieve true immersion with audio. We can currently use various video-audio teleconferencing tools to achieve a sense of connectedness over distances, and the ability to convey humanness and emotion has increased with current tools. There are, however, two major sound-related challenges with current tools: (1) the **challenge of distance**, and (2) the **challenge of noisy environments**. The communication experience that people have with the current videoconferencing systems (e.g., Zoom) remains fundamentally different from in-person experiences. Using remote work and communication tools, participants experience issues such as speaking over each other, not understanding speech because of surrounding noise, and lack of fluidity in conversations. Human communication systems need to be better, and Dr. Mehra and his team aim to create virtual sounds that are perceptually indistinguishable from reality, and to redefine human hearing in such environments. Their efforts are focused on two new capabilities: audio presence and face-to-face hearing in noisy environments.

It is important in virtual environments to achieve **audio presence** and provide a feeling that the source of a sound is in the same physical space as the person listening to the sound; a person who closes his/her eyes in such an environment should be able to feel that another person is next to him/her. To create audio presence, virtual sounds must be created with extremely high fidelity and be indistinguishable from real-world sounds to achieve the same level of communication as in real life. One challenge is incorporating **individual hearing characteristics**, represented via a *head-related transfer function* (HRTF). A person's individual hearing characteristics come from the shape of his or her ears, which differ from person to person and make sound interactions very personalized. Further, since sound, bouncing off from different surfaces, can reach an ear at different times and at different volumes, it is truly challenging to account for each ear when aiming to create audio presence. Another challenge is to account for the directionality of the sound and the impact of the individual person's body on the sense of a sound's direction. To achieve audio presence, we must enable **spatial audio** and sound directionality to better mimic how sound works in real life. In essence, we need the ability to define each person's personal HRTF. How do we generate someone's HRTF without physical measurements? How do we predict HRTF from simple input like photos or videos? The characteristics of a **room's acoustics** pose another challenge to achieving audio presence. If we can correctly calculate the acoustics of a space, and spatialization of audio, we can get much closer to sounds that are indistinguishable from reality.

Immersive audio must solve the problem of face-to-face verbal communications in noisy environments. A person should be able to **control the soundscape** (the social, psychological, and aesthetic quality of the acoustic environment), including determining what one wants to hear, and suppressing what one does not want to hear, given the soundscape of the location where the person is present. Future virtual reality (VR) glasses should be able to adjust to each person's personalized hearing ability and optimize the person's ability to hear. Achieving such capabilities requires beamforming, deep learning algorithms, and advanced noise-cancellation methods and devices. Sophisticated next-generation headphones will need to be able to infer attention intent, especially with respect to knowing what sound sources to enhance, sound enhancement using beamforming, and advanced noise cancellation.

### 6.2. Case for End-to-End Hardware-Software Co-Design for Immersive Systems

To achieve the full potential of immersive/XR computing, major advances in augmented, virtual, and mixed reality (AR/VR/MR) hardware, software, systems, and algorithms must occur. There is still an order-of-magnitude gap in performance, power, and user quality of experience between current systems and what is desired for future systems. For example, the compute power budget for comfortable, all-day-wearable AR glasses is about 200 mW, but current state-of-the-art systems expend several watts.

*Dr. Sarita Adve (UIUC)* made the case that with the end of Moore's law and Dennard scaling, bridging that gap will require research in designing end-to-end quality-of-experience-driven systems based on hardware-software-algorithm codesign. Further, given the stringent latency and bandwidth demands and the promise of multiuser XR applications, this co-design will have to encompass multiple end-user devices, the edge, and the cloud. She identified several challenges to performing such research: (1) the large diversity of expertise needed to develop XR systems, including graphics, video, audio, optics, and haptics; (2) the lack of disciplined co-design methodologies that span hardware, compilers, runtimes, networks, and algorithms from a multitude of domains; (3) complex and as-yet evolving metrics to measure the end-to-end goodness of XR systems; and (4) the lack of open-source reference XR systems and benchmarks that can enable such end-to-end systems research.

Dr. Adve presented an XR testbed called **ILLIXR** (Illinois Extended Reality testbed)[16], an open-source, end-to-end XR system designed to address the above challenges. ILLIXR supports XR perception, visual, and audio subsystems consisting of state-of-the-art sensors and components (e.g., visual inertial odometry, scene reconstruction, asynchronous reprojection, and 3D spatial audio encoding and decoding), all orchestrated through a flexible and efficient runtime system. ILLIXR runs XR applications that conform to the OpenXR interface, an emerging standard for XR applications. It runs on Linux PCs and embedded systems (e.g., NVIDIA Jetson), provides the option to offload some components to the cloud (e.g., AWS), and displays images on multiple commercial headsets. It provides extensive telemetry, enabling extensive power, performance, and quality-of-service measurements and insights on a fully functional XR system.

ILLIXR has led to a consortium with industrial and academic partners with the goal of democratizing XR research, development, and benchmarking. The consortium aims to establish a reference open-source testbed, a standard benchmarking methodology, and a cross-disciplinary R&D community for XR systems. ILLIXR is already being used in a variety of research projects to enable advances in XR hardware, software, systems, and algorithms, with the goal of improved end-to-end user quality of experience. The research includes design of new hardware accelerators that are co-designed with innovations in software algorithms for XR; co-design of 2.5D and 3D packaging technologies (for sensors, compute, and memory) with architectures and algorithms; techniques for offloading XR components from power-constrained wearables to edge and cloud servers; compilation technologies; scheduling and runtime system design; quality metrics; and more.

Achievement of the full potential of immersive computing will require continued investment in such end-to-end quality-of-service-driven, hardware-software-algorithms co-designed research, and open-source hardware and software infrastructure projects that can democratize such research.

### 6.3. Understanding the Brain to Enable Immersion

A big and complex question for immersive computing is, if we understood how the brain functions and could interpret brain activity, could we then identify the tasks a person wants to perform, and thus better immerse the person in his/her virtual environment seamlessly as he/she completes tasks? This is a difficult question with multiple facets. One is the ability to measure brain activity via a brain interface and have the data analysis capabilities to identify tasks and intent based on the data gathered on the activity. *Dr. Sanmi Koyejo (UIUC)* talked about **measurements of brain activities** and discussed our current understanding of the brain, and the role artificial intelligence (AI) can play in interpreting brain activity data to determine what tasks a person intends to undertake.

Measurements of brain activity have come a long way from the pseudoscience of phrenology to today's use of magnetic resonance (MR). Advances in MR were initially enabled by low-power magnets and then by high-magnetic-power imaging. Those brain interface instruments have enabled us to gain brain images and use them for scientific and medical purposes. Now that we have measurements of the brain obtained via **brain images**, researchers are seeking answers to questions on whether we can predict human activity from the brain, and whether we can go from observing brain activities to figuring out what the brain is doing. It is known in the brain imaging community that the interpretation of brain activity is tied to how **concepts** are encoded in the brain. Much work has been done and continues to be done in determining **relationships between concepts and brain regions**. For example, it is well known that many brain functions are primarily performed in either the left or the right side of the brain. Dr. Koyejo studies probabilistic graphical models for brain data to find such relationships. His results show, for example, that brain activity generates brain images corresponding to individual words or phrases. The next step is to identify which brain regions are associated with which tasks. That will require new deep-learning architectures that exploit spatial structure.

---
[16] https://illixr.org/

Several challenges will be involved in answering the above questions.

- New deep learning architectures are needed, but a challenge is to find a way to learn over **small data sample sizes**, since brain imaging and the corresponding generation of large sets of data on brain imaging samples are expensive. One must address issues such as variability and overfitting. Another issue is the mapping of millions of signals to a small number of possible activities.
- From the challenge of small data sample sizes follows the challenge of **synthetic data generation**, or exploration of data augmentation for machine learning algorithms. A question to be answered is, can we synthesize power curves for new experiments?
- Another challenge in understanding brain activity is to understand higher-order behavior in the brain, i.e., **dynamic brain connectivity**. We cannot measure the dynamics directly through brain imaging. To make progress in our understanding of higher-order behavior, we need (1) estimation techniques for dynamic networks, (2) a greater understanding of the topology of brain networks, and (3) insights on how dynamic networks drive human behavior.

The above challenges clearly indicate that we need major engineering innovations and AI advances to drive the future of brain imaging analysis. We need advances in simultaneous multimodal data collection that includes MRI (magnetic resonance imaging), EEG (electroencephalography), MEG (magnetoencephalography), and ECOG (electrocorticography), among others. Further, algorithmic advances that can accelerate the work include distributed and hierarchical machine learning algorithms, such as federated learning algorithms, that can work over distributed datasets. We need new ways to probe brain activity, new ways to model brain activity, and new statistical applications. These challenges will require advances not only from engineering disciplines but also from psychology and cognitive science. Psychology and cognitive science researchers have improved the classification of simple brain tasks, but we need to go beyond simple tasks. To move in that direction, open datasets and open science would be beneficial to our efforts to understand brain activity and, ultimately, enable immersion in virtual environments based on brain activity and task prediction.

### 6.4. Panel Discussion

The panel discussed topics related to (1) immersive audio work at academic institutions, (2) advanced XR systems and the corresponding hardware-software co-design, and (3) the role of neuroscience in advancing understanding of the dynamic network structure of the brain.

**Immersive Audio Research in Academic Environments**: Academic environments have an important role in research on immersive audio since academic researchers can study problems under controlled conditions. This allows the exploration of bleeding-edge cases, such as creation of **sounds indistinguishable from reality** or improvement of control in changing sounds according to personal user needs. Second, immersive audio requires solutions that can be adapted to a large problem space. Some sub-problems in the problem space are being and will be addressed by academia, such as **beamforming**. Most current beamforming solutions work only on static devices, but this needs to change for AR/VR devices, as they reside on a user's head and are thus constantly moving. Third, machine learning algorithms work well on recorded audio and video datasets, but not on data coming from **moving arrays**. New frontiers must be explored, and academia can lead industry here.

**Advanced XR Systems**: AR/VR will be the next interface to computer systems. To make that a reality, integrated multidisciplinary research must be undertaken. Problem-solving in terms of performance efficiency in one part of the system stack (e.g., hardware and compilers), without consideration of impacts on other parts of the stack, will not be adequate. Different domains must collaborate if new plug-and-play chips and systems (both physical and simulated) are to be developed to move us to scalable end-to-end XR solutions.

**Relation of Neuroscience, AI, and Dynamic Structures of the Brain**: The success of AI was inspired by early breakthroughs in neuroscience forty years ago. The discovery of dynamic network structures in the brain created great excitement at the time. We must now build on such prior knowledge and recent innovations and discoveries to develop learning algorithms that are more efficient. Current results are mixed. We understand that the human brain is much faster at inference tasks (e.g., recognition of an object) than ML models. The brain can also deal much faster with the nuances of environments and in tolerating errors (even if it is also prone to bias). There is thus a significant gap between what ML tools can do and what the brain can do. The hope is that findings from neuroscience will help drive new ML tools to bridge the gap.

## 7. Robotics and Autonomy

Future computing paradigms must include concepts, algorithms, and modeling of autonomous systems, including hardware and software that can be deployed within robotic systems in different domains. Such domains include transportation (including self-driving cars, robo-taxis, and autonomous trucks), manufacturing (including autonomous assembly), and agriculture (including autonomous combines, weed-pulling robots, autonomous crop inspection, and fruit-picking systems). In this session, three topics were discussed: (1) the level of autonomy in self-driving cars such as robo-taxis; (2) the level of autonomy required in human-robot systems when deployed within transportation, manufacturing, or agriculture; and (3) socially aware autonomy for multi-robot interactions.

### 7.1. Level of Autonomy for Self-Driving Cars

Prior to 2020 (pre-COVID times), Americans spent fifty billion hours driving per year. Accidents were, and still are now, too common and highly undesirable. *Dr. Daniela Rus (MIT)* studies this problem and asks questions such as, what if cars could learn how to drive to avoid collisions? What if we could make riding in a car fun for the passengers? Passengers could read books while their car is in autopilot mode, or the car might even take the initiative to do chores on its own at night, such as getting gas. Dr. Rus used her talk to explain why this is still a long way away.

Autonomous driving has been a dream for a long time. The first autonomous coast-to-coast drive across the U.S. occurred in 1995; a NavLab (CMU Navigation Laboratory)[17] team drove the car in autonomous mode except for when the car needed to take exits or found itself in a difficult situation. In those situations, the human driver took over. The world's first autonomous highway drive occurred in 1986, when Ernst Dickmann[18] had a car drive autonomously on an empty part of the Bavarian Autobahn. Dynamic computer vision contributed to the car's ability to drive autonomously. Solutions were neither robust nor fast: it took 10 minutes to analyze each image, whereas an autopilot system requires analysis of at least ten frames per second to operate a car at a reasonable speed. Image processing has improved significantly since then, and driverless cars rely on computer vision that can analyze one hundred frames per second.

New questions and key parameters are emerging. How complex is the environment? How complex are the anticipated car-environment interactions? A paramount question is: How will the car cope with **uncertainties**? Sensors do not work well in bad weather, i.e., heavy rain and snow. Uncertainties increase with congestion and with erratic driving of nearby human drivers, and unexpected obstacles can be encountered, such as cows crossing a road. How do we give self-driving vehicles the ability to operate safely on roads with both self-driving and human-driven vehicles? It will be extremely challenging to transform the autonomous driving pipeline of today's classical autonomous vehicle, consisting of a robot platform extended with sensors such as lidar and cameras, to incorporate perception, estimation, and learning.

Environmental complexity gives rise to additional challenges, such as the need for learning within the autonomous controller of the self-driving car. Can we use a large dataset to learn how humans drive in similar situations? Can we obtain datasets that cover a sufficiently wide range of scenarios? We need **end-to-end learning systems**. Dr. Rus and her team have developed such an end-to-end learning system, executing pipeline of tasks ranging from the capture of raw input (pixels from the camera and rough topological street maps), to object and environment recognition, to vehicle trajectory inference, and to inference of a continuous probability distribution of inferred trajectories of the vehicle. The end-to-end learning framework uses deep neural network models to process both raw sensor data and additional map information to allow the vehicle to navigate point-to-point. This approach enables **uncertainty-aware predictions**, including generalization capabilities for the vehicle when it is driving on unfamiliar roads and through unfamiliar intersections, e.g., driving safely on roundabouts even if it had no prior training data on roundabouts.

Prior research on self-driving cars shows that one must design **systems that are robust** to unexpected events. For example, a sensor failure could cause a crash with a vanilla end-to-end algorithm, but with uncertainty-aware control, the car would be able to handle sensor failures autonomously. Another aspect of robustness is the need to obtain training data for new domains and edge cases. Such data are difficult to obtain for many real situations, and training data from other situations are difficult to transfer to new domains and edge cases. Dr. Rus and her team developed a platform, called Vista, which is a **photorealistic, data-driven simulation engine**. Vista allows the simulation of real-world situations and navigation within an envelope without human supervision. Vista delivers a transferable reinforcement learning framework with a fully connected neural network for analyzing images, processing motor neurons, and delivering control to steer the vehicle. One of the concerns coming out this research is the need for sustainability. Traditional neural networks need thousands of neurons to converge to answers, consuming large amounts of energy. Deep neural networks have a large carbon footprint. One approach presented in this talk is to develop new models for artificial neurons and architectures for wiring the neurons. The proposed neural system for self-driving cars' steering problems uses *neural circuit policies* (NCP), which make it possible to wire learning architectures inspired by nature and map them into a **small number of neurons**, **i.e., only nineteen neurons**, not the 100,000 simple ones utilized before.

---

[17] https://www.cs.cmu.edu/afs/cs/project/alv/www/
[18] https://www.politico.eu/article/delf-driving-car-born-1986-ernst-dickmanns-mercedes/

The next challenge to consider is how to enable autonomy of self-driving cars in hybrid environments in which robot cars must interact with human-driven cars. In these environments, one must know signage rules, interpret silent interactions between people and vehicles, and understand the overall complex road behavior logic. This gives rise to several questions, such as what happens if a human and a robot car do not understand each other, or what happens if robot cars get confused about whether a human is yielding. Such mixed environments give rise to situations that emerge from social dilemmas. One potential solution is to introduce the concept of **Social Value Orientation (SVO) Rings** that capture personalities and human preferences in such situations. For example, some human drivers could be either prosocial and altruistic, or aggressive and egoistical, in each situation. If one could measure SVOs, the self-driving car could have rules such as "slow down and wait if incoming traffic is egoistic." One could then model this situation as a "best response game," aim for Nash equilibrium, and calibrate rewards on real data sets. Studies have shown that SVOs improve trajectory prediction by 25% and improve safety.

Self-driving cars have a long way to go before they will be legal on most roads. There are several levels of autonomy. To enable the highest autonomy level, Level 5, for self-driving cars, new policies must be instituted, and technical challenges must be overcome. A fundamental question from the policymaking side is to determine what a self-driving car should and shouldn't do, and what can it do. The future of autonomy must deal with technical problems such as the presence of many sources of uncertainty, neural circuit policies for reducing uncertainty, and SVOs for decision-making and control, among many others. Given all the remaining barriers to the achievement of self-driving robo-taxis, it's unclear which will come first: Level 5 autonomy or flying cars.

### 7.2. Structure of Human-Robot Systems

As discussed above, robotic autonomous systems will need to operate safely in complex environments in which humans and robots interact, work together, cooperate, and rely on each other. *Dr. Katie Driggs-Campbell (UIUC)* studies the question of safety of autonomous systems in complex human-robot environments within domains such as transportation, manufacturing, and agriculture. These domains represent highly interdisciplinary environments. Fundamental questions in computing arise from domain-specific grand challenges.

In modern manufacturing plants, there is a great deal of **control** over the manufacturing environment; this domain has very well-defined processes. Thus, there is less uncertainty in the context of human-robot structures than in, for example, transportation. In transportation, there are road rules, but much more **uncertainty** in the human-robot structure, as discussed in Section 7.1. The agricultural domain is the most dynamic of the three environments. The environment is constantly changing with weather, the variable growth of plants, and the presence of not only humans but animals. Technological services important to operations, such as GPS, might be absent, or available only intermittently.

The structure of human-robot systems is greatly influenced by the **domain** in which they are deployed, by various **failure characteristics**, and by assumptions about the viable **safety tools** that must be considered. For example, in the case of agriculture, if a failure occurs, it can be difficult even to identify the cause of the failure; one must have extensive anomaly detection capabilities present. In transportation, if self-driving cars get into complex environments, safety correspondingly becomes very complex. Simulation-based validation in such situations must be used to show safety prior to any road tests. In manufacturing, safety approaches can utilize model-based assessments, which will check carefully defined specifications and constraints that are known to the designer.

Another challenge in human-robot systems emerges from the **context of systems**. None of the robots operate alone. They will always be interfacing with people in some way. Hence, the structure of human-robot systems influences how we codify the level of autonomy and modes of interactions. For example, in transportation, it is useful to consider a hybrid mode of operation in which an autonomous car system handles most of the driving, while a human monitors it and takes over in difficult situations. In manufacturing, we have highly automated systems for which many tasks are executed by robots. But there are also collaborative manufacturing environments in which people and robotic systems work closely together and have tight coupling. The system context in agriculture is vastly different. In this domain, the human-robot interaction might be loose. The human might take on the role of a monitor and controller. However, as the level of autonomy increases, the human role might become one of one-to-many control, and full remote supervision might eventually be possible, such that one human will take on the role of a supervisor and monitor all the robots in an entire farm. Full robot autonomy is unlikely, however. Given the tremendous uncertainties in agricultural environments, it is very unlikely that there will be no human in the loop.

Overall, there are different **levels of autonomy**, as alluded to in Section 7.1. We are currently at Levels 1 to 4 in the agriculture and transportation domains, because it is possible to deploy full autonomy in some of these environments, some of the time. A lot of research is needed if we are to increase the levels of automation in the various domains towards Level 5.

### 7.3. Socially Aware Autonomy: Planning and Control of Multi-Agent Interactions

When robots want to collaborate and interact with each other and humans, the following questions come up. What are the planning steps towards achieving seamless interactions of robots in multi-agent domains? And what are the social norms to which these robots should adhere? *Dr. Negar Mehr (UIUC)* studies multi-agent interactions, especially with respect to control and planning challenges. Many research topics must be addressed in this area. They include robot dependency on all agents' decisions, and systems that would enable accountability for this dependence among multiple agents.

Individual robots are usually programmed to minimize their cost functions in executing certain tasks. In the case of multi-agent interactions among robots and/or humans, individualistic control does not work, because the action of each agent affects the other agents' actions. Interactions among multiple agents and the resulting couplings among agents' decisions are studied via **game theory**. The goal is to reach equilibrium, wherein each agent makes the best decision once the policy of all the other agents has been fixed. To achieve the best response, each agent must solve a **coupled optimization problem**. This is a powerful theoretical framework. Finding equilibrium via this framework in **real time**, however, is difficult, even for simple games. To solve the problem in real time, new solution methods must be sought.

Dr. Mehr studies the decision-planning and control frameworks for multi-agent interactions at three levels of interaction and decision-making. First, one must consider **interactive motion planning**, wherein each robot solves a set of coupled motion-planning problems in real time, and in a scalable and efficient fashion. Interactive motion planning belongs to the class of **potential games.** A *potential differential game* is one in which equilibria can be found efficiently and in a scalable manner. Second, one must consider **interactive task planning**, in which higher-level decisions must be made. Examples of questions that require higher-level decisions might include "Who will do what to succeed?" and "Who will take what role?" A good example of a higher-level decision that robots must take on is the following. Let us assume that two robots must transport heavy objects together to a desired position. If two humans were to move similar objects, one person would usually follow the other person. In the case of robots, if role assignment is not done properly, the robots may collide and fail to accomplish the task. Interactive task planning belongs to the **class of meta games**. It has been shown that agents can accomplish a task by frequently switching roles. One efficient algorithm is the Empirical Game-Theoretical Analysis (EGTA) algorithm developed by Dr. Mehr and her team. Third, one must consider **societal-scale interactions** and the context of the domain in which the agents are embedded. For example, in a transportation context, such as a fleet of autonomous cars sharing a road with human-driven cars, we must ensure a societal benefit. There is a major paradox, called **Braess's paradox**[19], regarding decision-making with respect to societal-scale interactions in transportation. Braess's paradox is the observation that adding one or more roads to a road network can slow down overall traffic flow through it, i.e., constructing new roads or increasing road capacity can increase the overall congestion. Are there similar paradoxes to be considered in mixed-autonomy environments where human-driven and autonomous cars coexist? A key question for societal-scale interactions is, how do we achieve altruistic autonomy?

Overall, one must think carefully about the coupling between agents' trajectories, planning for agents' high-level choices, and accounting for societal implications of their interactions.

### 7.4. Panel Discussion

The panel discussion concentrated on three topics: (1) modeling of human behavior, (2) expert driver-based models, and (3) consideration of NCP (neuron circuit policy) chips as an alternative to GPU chips to allow machine learning on edge devices.

**Modeling of Human Behavior**: It is very important to have extensive human behavior models for self-driving cars, especially in an urban environment. We currently have two types of solutions: (1) model-based solutions, which are nice in that we can analyze them and get guarantees but are not reliable or scalable; and (2) data-driven solutions, which tend to capture nuanced trends of human motion and raise the level of abstraction when we're thinking about the problem, but are prone to mistakes. The best approach is to have both: reasoning in higher levels of abstraction, and capturing of and responding to mistakes in real time.

Interactions between humans and self-driving cars remain one of the most challenging aspects of autonomous driving. For example, consider a situation in which a person pops out from nowhere in front of a vehicle. What would be the interaction between an autonomous vehicle and the pedestrian in such a situation? Today's autonomy solutions are not ready for complex environments. There is way too much uncertainty. What is possible today is **driver-assisted systems** that can see beyond what the driver sees. Extension of perceptual abilities, e.g., to provide warning signals, is already in place, and more is coming.

---

[19] https://en.wikipedia.org/wiki/Braess%27s_paradox

Mathematical models of personalities are also available. We can detect people's **personalities**, but we do not have many models for other aspects of human behavior. Research is being conducted on behavior predictions. Large sets of individual drivers' behavior data are being collected and analyzed, e.g., for identification of distracted drivers, but it is not obvious what kinds of models and insights one can obtain and how much of the data is translatable and generalizable. Essentially, each driver behaves differently, and aiming to have personalized models for every human driver is not practical at scale.

**Expert-driver-based Models**: Instead of deriving individual human behavior models for use in planning and control frameworks, one could capture and learn directly from expert human drivers and codify data for self-driving cars based on expert drivers. There is no shortage of real-world data for autonomous driving, which has led to reliable predictions most of the time. However, there is still a challenge in capturing long-tail events and rare situations, both in terms of representation in data and in determining what the self-driving car should do.

**NCP Chips**: The neuron circuit policy (NCP) method introduced a significant reduction in neural network complexity. It would be interesting to bring this method to hardware and develop a new NCP chip along the lines of other neural network hardware accelerators for large neural networks. Doing so is not possible at this point, but there is potential for the future. One must consider that in the current deep neural networks, all neurons are identical, and each one computes a simple step function and does one kind of estimation. NCP requires a different model of the neuron. In NCP, neurons are not all equal; they are specialized. That approach works for tasks with time series data (e.g., steering data), but not for object detection and recognition tasks. It may take time to develop NCP chips that could be an alternative to GPU chips in executing machine learning at edge devices.

8. **Closing Remarks**

The **Future of Computing Symposium** brought together UIUC and external researchers to discuss research topics that merit attention for their potential to move current computing paradigms forward. Pursuing such research is important to ensure that computing offers solutions to many of the technological, environmental, and societal challenges humanity faces. This symposium, held as part of CSL's 70$^{th}$ anniversary celebrations and with many of its junior and senior members in attendance, presented many hot topics in computing.

As discussed throughout this document, finding solutions to computing challenges goes hand-in-hand with ethical considerations, and we recommend that interested readers look at the materials from the **Symposium on Artificial Intelligence and Social Responsibility** (https://csl.illinois.edu/events/ai-social-symposium), another event held in celebration of the 70$^{th}$ anniversary of the Coordinated Science Laboratory. That symposium presented numerous perspectives on social responsibility with respect to AI technologies and computing.

**Acknowledgement**: The authors would like to thank Jenny Applequist and Dr. Normand Paquin for their assistance during the symposium and the white paper preparation via editing and constructive comments.